\documentstyle[12pt]{article}
\begin{document}
\renewcommand{\theequation}{\arabic{section}.\arabic{equation}}
\begin{titlepage}
\vspace*{1cm}
\begin{center}
{\LARGE \bf             Quasiclassical Approach to Two-level\\
\vspace*{0.2cm}
                        Systems With Dissipation}
\end{center}
\vspace*{0.5cm}
\begin{center}
{\large \bf            Mo-Lin Ge and Lei Wang} \\
{\bf                   Theoretical Physics Division,
                       Nankai Institute of Mathematics,\\
                       Tianjin, 300071, People's Republic of China}
\end{center}
\vspace*{0.4in}
\begin{abstract}
\baselineskip=23pt
The quantum dynamics of two-level systems under classical oscillator heat
bath is mapped to the classical one of a charged particle under harmonic
oscillator potential plus a magnetic field in a plane. The behavior of
eigenstates and tunneling and localization are studied in detail. The
broken symmetry condition and Langevin-like dissipative equation of
motion are obtained. Some special dynamic features are considered.
\end{abstract}
\vspace*{4.5cm}
\end{titlepage}
\newpage
\baselineskip=23pt
\section{Introduction}
\setcounter{equation}{0}
Tunneling and localization in dissipative system have attracted much
attention for their theoretical and experimental studies have much
enriched our understanding on the real physical world [1-12].
For macroscopic and mesoscopic quantum tunneling phenomena two classes of
processes are interested: the decay from metastable states and quantum
coherence tunneling. The former had been extensively investigated by
many authors \cite{Ca1,Ha1}. The studies indicate that the tunneling
probability
is significantly suppressed at strong dissipation if we require the
dissipative Langevin equation
\begin{equation}
M\ddot{q}+\eta\dot{q}+\frac{dV}{dq}=F_{ext} \label{e0}
\end{equation}
to be valid in semiclassical regime. The latter is a more difficult problem.
A remarkable two-level model was
given by Leggett {\em et al.} \cite{Ca1,Le1,Ch1}, and
the possibility of localization in the presence of dissipation was
intensively studied. However, it is difficult to give a intuitive
and uniformed description to shed light the interaction between
two-level system and environment. In this paper, we shall study the
behavior of two-level model
of Leggett {\em et al.} by reducing the quantum dynamics
of the two-level system to the movement of a charged particle in a harmonic
oscillator potential under a magnetic field in a plane through a mapping.
The method is the generalization of that discussed in ref. \cite{Wa1}. In this
approach the quasiclassical dissipative dynamics for two-level
quantum evolution is mapped to a classical mechanical problem.
Here ``quasiclassical'' means that we shall treat the environment
oscillators classically and the two-level system interacted with
the environment oscillators by quantum mechanics.

Our main conclusion is the following:\\
(1) When the acceleration of environment oscillators is small, the total
energy values(system plus environment) $E_c< -\hbar\omega_0$
as increasing of the interaction between the system and environment.
The states are localized in the left or right of a symmetric double-well.\\
(2) The solution is stable.\\
(3) The dynamic equation for $S(t)=\langle  \sigma_3\rangle$
is obtained that explores
the main physical picture of the system interacted with environment.

\section{Ground State}
\setcounter{equation}{0}

The Hamiltonian for spin-boson is given by \cite{Le1}
\begin{equation}
\hat{H}=-\frac{1}{2}\hbar\Delta\sigma_{1}+\frac{1}{2}
\sum_{j=1}^{N}(\frac{p^2_j}{m_j}+m_j\omega^2_jx^2_j)+\frac{q_0}2\sigma_3
\sum_{j=1}^{N}c_jx_j \label{e1},
\end{equation}
where $\Delta=2\omega_0$ is the energy splitting between the two levels,
$\sigma_{1},\sigma_{3}$ are Pauli matrices and $x_{i},
p_{i}$ are the coordinate and momentum of $i$-th oscillators, respectively.
Eq~(\ref{e1}) can be recast to the form
\begin{equation}
\hat{H}=-\frac{1}{2}\hbar\Delta\sigma_{1}+\frac{q_0}{2}\sigma_{3}
\sum_{j=1}^{N}c_{j}(\frac{\hbar}{2m\omega_j})^{1/2}(a_j+a^{+}_j)+H_R
\label{e2},
\end{equation}
where $H_R=\sum^N_{j=1}(a^{+}_ja_j+\frac 12)\omega_j\hbar$, and
\begin{equation}
[a_j, a^+_i]=\delta_{ij},\ \ a_j=\frac 1{\sqrt{2}}(\sqrt{\frac{m\omega}{\hbar}}
x_j+i\frac 1{\sqrt{m\omega \hbar}}p_j),
\end{equation}
The wave function obeys
\begin{equation}
i\hbar\frac{\partial \Psi}{\partial t}=H\Psi.
\end{equation}
Making the transformation
\begin{equation}
\Psi^{\prime}=e^{-\frac i{\hbar}H_Rt}\Psi,
\end{equation}
one obtains
\begin{equation}
i\hbar\frac{\partial \Psi^{\prime}}{\partial t}
=\left\{ -\frac{1}{2}\hbar\Delta\sigma_{1}
  +\frac{q_0}{2}\sigma_{3}\sum_{j=1}^{N}c_{j}(
  e^{i\hbar\omega_jt} a^+_j+e^{-i\hbar\omega_jt}a_j)\right\}\Psi^{\prime}.
\end{equation}
Obviously the non-commutativity between $x_i$ and $H_R$ only gives rise to
the canonical transformation by $b^+_j=e^{i\omega_jt}a^+_j$ that preserves
$H_R=\sum^N_{j=1}(b^+_jb_j+\frac 12)\omega_j\hbar$. Keeping this canonical
transformation in mind and still denoting the transformed coordinates by
$x_j$ we have
\begin{equation}
i\hbar\frac{\partial \Psi^{\prime}}{\partial t} =\left\{
-\frac{1}{2}\hbar\Delta
\sigma_{1}+\frac{q_0}{2}\sigma_{3}\sum_{j=1}^{N}c_{j}x_j)\right\}\Psi^{\prime}.
\end{equation}
Making further transformation
\begin{equation}
\Psi^{\prime}=exp\{-i\xi (t)\sigma_3\}\phi
\end{equation}
where
\begin{equation}
\xi (t)=\frac{q_0}{2\hbar}\int^t_0\sum_ic_ix_i(\tau)d\tau=\frac{q_0}{2\hbar}
   \int^t_0d\tau X(\tau),
\end{equation}
we have
\begin{equation}
i\hbar\frac{\partial \phi}{\partial t} =(-\frac{\Delta}2)
  (\sigma_+e^{iq_0\xi}+\sigma_-e^{-iq_0\xi})\phi
\end{equation}
or
\begin{eqnarray}
i\frac{\partial \alpha}{\partial t}
&=& (-\frac{\Delta}2)e^{iq_0\xi}\beta\nonumber,\\
i\frac{\partial \beta}{\partial t} &=& (-\frac{\Delta}2)e^{-iq_0\xi}\alpha
\label{e4},
\end{eqnarray}
where $\phi=\left(\begin{array}{c}\alpha\\ \beta\end{array}\right)$. The
normalization condition is
\begin{equation}
|\alpha|^2+|\beta|^2=1 \label{e5}.
\end{equation}
Defining planar vectors
\begin{eqnarray}
\vec r &=& \mbox{Re}\alpha\vec{e_x}+\mbox{Im}\alpha\vec{e_y},\ \
    \alpha=r e^{-i\theta}\nonumber,\\
\vec\rho &=& \mbox{Re}\beta\vec{e_x}+\mbox{Im}\beta\vec{e_y},\ \
    \beta=\rho e^{-i\varphi}
\end{eqnarray}
and
\begin{equation}
\vec B(t)=\frac{q_o}{\hbar}(\sum_jc_jx_j)\vec{e_z}=\sum_jc_j\vec{B_j}.
\end{equation}
Eq.~(\ref{e4}) can be recast to the form:
\begin{eqnarray}
\frac{d^{2}}{dt^{2}}\vec{r} &=& -\omega_{0}^{2}\vec{r}
   -\frac{d\vec{r}}{dt}\times\vec{B} \label{e6},\\
\frac{d^{2}}{dt^{2}}\vec{\rho} &=& -\omega_{0}^{2}\vec{\rho}+
    \frac{d\vec{\rho}}{dt}\times\vec{B} \label{e7}.
\end{eqnarray}

Henceforth the environment
is viewed to be classical, namely any oscillator
in the heat bath is treated as one of the driven harmonic oscillator
with the driven force proportional to the time-dependent average value of
$\sigma_z$ through the canonical equation of motion:
\begin{equation}
\frac{d^{2}}{dt^{2}}B_{i}+\omega_{i}^{2}B_{i}+\frac{q_{0}^{2}c_{i}^{2}}
{2m_{i}\hbar}S(t)=0 \label{e8}
\end{equation}
where
\begin{equation}
S(t)=\langle  \sigma_{z}\rangle=\langle  \Psi|\sigma_3|\Psi\rangle
 =\langle  \phi|\sigma_3|\phi\rangle.
\end{equation}
Because of eqs.~(\ref{e5}), it is easy to know that
\begin{eqnarray}
&& S(t) = 2r^2-1=1-2\rho^2 \label{e13},\\
&& \langle  \sigma_1\rangle = r^2 \dot{\theta}, \\
&& \dot{\vec{r}}\ ^2 = \omega^2_0\vec{\rho}^2,\ \
\dot{\vec{\rho}}\ ^2=\omega^2_0\vec{r}\ ^2 .
\end{eqnarray}
The energy conservation reads
\begin{eqnarray}
\frac 12\dot{\vec{r}}\ ^2+\frac 12 \omega^2_0\vec{r}\ ^2
&=& \frac 12\omega^2 \label{e14},\\
\frac 12\dot{\vec{\rho}}\ ^2+\frac 12 \omega^2_0\vec{\rho}\ ^2
&=& \frac 12\omega^2.
\end{eqnarray}
The total energy is
\begin{eqnarray}
E_{\mbox{total}} &=& \langle  \Psi|H|\Psi\rangle\nonumber\\
&=&-\hbar r^{2}\dot{\theta}+q_{0}\hbar(r^{2}-\frac{1}{2})
\sum_{i=1}^{N}c_{i}x_{i}+\sum_{i=1}^{N}\frac{m_{i}}{2}(\dot{x}_{i}^{2}
+\omega_{i}^{2}x_{i}^{2}).
\end{eqnarray}
Noting that the energy $-\omega_0\hbar\langle  \sigma_1\rangle$
is precisely the minus sign of
angular momentum of charged particle moving in a applied magnetic field
$\vec B(t)$ shown by eq.~(\ref{e6}).

Now the two-level system obeys eq.~(\ref{e6}) or (\ref{e7}) which looks
like the dynamics of charged particle experiencing harmonic force in a
magnetic field perpendicular to the plane in which the particle moves.
This picture is similar to Feynman-Vernon-Hellwavth's Bloch vector
stratagem \cite{Fe1}. The dynamics of oscillators in heat bath is
described by eq.~(\ref{e8}).

Eqs.~(\ref{e6}) and (\ref{e8}) allows to describe the shift of ground state
of two-level system when the interaction between the system and environment
is getting large.

Looking at eq.~(\ref{e6}), we see
that the circular movement orbit of charged particle
can be verified as the ground state when $\ddot B_i$ does not vary with time
at $\vec{r}=\vec{R}$:
\begin{equation}
B=\sum_jB_j=\left(\sum_j\frac{q^2_0c^2_j}{2\hbar m_j\omega^2_j}\right)(1-2R^2)
\label{e9}.
\end{equation}
Denoting $\vec R=R\vec e_r$, $\dot{\vec r}|_{\vec r=\vec R}=
v\vec e_{\theta}$, we obtain from eqs.~(\ref{e9}) and (\ref{e6})
\begin{equation}
(2R^{2}-1)(1-\frac{q^{2}_{0}}{2\hbar\omega_{0}}vR\sum_{i=1}^{N}
\frac{c_{i}^{2}}{m_{i}\omega_{i}^{2}})=0 \label{e10}.
\end{equation}
It is easy to know that
\begin{equation}
vR=\omega_0R\sqrt{1-R^2}=\frac{2\hbar\omega^2_0}{q^2_0}\left[
\sum_j\frac{c^2_j}{m_j\omega^2_j}\right]^{-1}.
\end{equation}
Hence eq.~(\ref{e10}) becomes
\begin{equation}
R^4-R^2+\frac{4\hbar^2\omega^2_0}{q^4_0}\left(\sum_j\frac{c^2_j}
{m_j\omega^2_j}\right)^{-2}=0 \label{e11}.
\end{equation}
The solution of eq.~(\ref{e11}) corresponds to the lowest-energy state of
$\phi$.

Denoting
\begin{equation}
D=1-\frac{16\hbar\omega^2_0}{q^4_0}\left(\sum_j\frac{c^2_j}
  {m_j\omega^2_j}\right)^{-2},
\end{equation}
we find\\
(1) for $D<0$, no solution.\\
(2) for $D=0$,
\begin{equation}
R=\frac{\sqrt{2}}{2},  v=\pm\frac{\sqrt{2}}{2}\omega_{0},  B=0
\label{e15}.
\end{equation}
(3) for $D>0$, two solutions:
\begin{equation}
R^2_{\pm}=\frac{1}{2}\left\{1\pm\sqrt{1-\frac{16\hbar^{2}\omega_{0}^{2}}
{q_{0}^{4}}
\left(\sum_{i=1}^{N}\frac{c_{i}^{2}}{m_{i}\omega_{i}^{2}}\right)^{-2}}
\right\}^{1/2} \label{e20}.
\end{equation}
In the case (2) $B=0$, $|\phi\rangle=\frac 1{\sqrt 2}e^{\pm i\omega_0t}\left(
\begin{array}{c}1\\ \pm 1\end{array}\right)$, the energy of system is just
the total energy(system plus invariant):
\begin{equation}
E_{\mbox{total}}=\langle  \Phi|i\hbar \frac{d}{dt}|\Phi\rangle=\mp \hbar
\omega_0
\label{e12}.
\end{equation}
It is obvious that for the absence of interaction between the system and
environment, i.e. the magnetic field vanishes, the system is nothing but
a bare two-level one, namely, the total energy system plus environment
is equal to $-\hbar \omega_0$, and the excitation state has the energy
$\hbar \omega_0$.

In the case (3) $D>0$ means that
\begin{equation}
\frac{q^2_o}4 \sum^N_{j=1}\frac{c^2_j}{m_j\omega^2_j}>\hbar\omega_0\label{e17}.
\end{equation}
Solutions of eq.~(\ref{e11}) reads
\begin{equation}
R_{\pm}=\left\{ \frac{1}{2}\left\{ 1\pm \left[1-\frac{16\hbar^{2}
   \omega_{0}^2}{q^4_0}\left(\sum_{i=1}^{N}\frac{c^2_i}{m_i
   \omega^2_i}\right)^{-2}\right]^{1/2} \right\}\right\}^{1/2}\label{e16},
\end{equation}
correspondingly,
\begin{equation}
B_{\pm}=\mp\frac{1}{2\hbar}\left\{ q_{0}^{4}(\sum_{i=1}^{N}\frac{c_{i}^{2}}
{m_{i}\omega_{i}^{2}})^{2}-16\hbar^{2}\omega_{0}^{2}\right\}^{1/2}
\end{equation}
which indicates that there exists the interaction between the system and
environment. Differing from eq.~(\ref{e12}) the total energy is
\begin{eqnarray}
E_{\mbox{total}} &=& -\hbar\omega_0\sqrt{\frac{1-R^2}{R^2}}+\frac{\hbar}2B
  +\frac 12\sum^N_jm_j\omega^2_jx^2_j\nonumber\\
&=& -\frac{q^2_0}8 \sum^N_{j=1}\frac{c^2_j}{m_j\omega^2_j}\left[1+\frac
  {16\hbar^2\omega^2_j}{q^4_0}\left(\sum^N_{j=1}\frac{c^2_j}
  {m_j\omega^2_j}\right)^2\right]< -\hbar\omega_0
\end{eqnarray}
with  degeneracy. Therefore, we find the new ground state with the
energy lower than $-\hbar\omega_0$. When the acceleration of $x_j$ can be
neglected, the physical picture is viewed as the following:
The two-level system has the ground state with $E_{\mbox{total}}=-\hbar\omega
_0$ and the excitation state with $E_{\mbox{total}}=+\hbar\omega_0$, if the
environment is completely ``frozen''. Whereas the ground state of the system
becomes double-degeneracy with energy lower than $-\hbar\omega_0$ as
increasing of interaction between the
system and environment that corresponds to the non-vanishing displacements
of the magnetic oscillators from their equilibrium points. We call the
degenerated states  localized states because in double-well system theory
they represent those states which mainly localized in the left or right
well with different parities. Later in this paper we shall show that the
bare two-level in the system becomes unstable and will decay into one of
the localized eigenstates with the energy lower than $-\hbar\omega_0$ in
the presence of dissipation.

\section{Stability}
\setcounter{equation}{0}

Eqs.(\ref{e6}), (\ref{e8}), (\ref{e13}) together with (\ref{e14})
form a set of equation determining both the quantum dynamics of the system
and semi-classical dynamics of oscillators in the bath. For $\ddot B_{i}$ can
be neglected the eq.~(\ref{e9}) holds. With this picture eq.~(\ref{e6})
takes the form:
\begin{eqnarray}
&&\ddot r-r \dot{\theta}^2+\omega^2_0r -\frac{q^2_0}{2\hbar}
    (2r^2-1)r \dot{\theta}\sum_j\frac{c^2_j}{m_j\omega^2_j}=0,\\
&&r \ddot{\theta}+2\dot r\dot{\theta}+\frac{q^2_0}{2\hbar}(2r^2-1)
    \dot r\sum_j\frac{c^2_j}{m_j\omega^2_j}=0\label{f1}.
\end{eqnarray}
Eq.(\ref{f1}) leads to
\begin{eqnarray}
&& r\dot{\theta}(r^4-r^2)J=L_0=\mbox{constant\ of\ motion}\label{f01},\\
&& J=\frac{q^2_0}{4\hbar}\sum_j\frac{c^2_j}{m_j\omega^2_j}.
\end{eqnarray}
$L_0$ is nothing but the effective angular momentum for the charged particle
moving in the magnetic field and
\begin{equation}
L_0=-E_0 \label{f20}
\end{equation}
where $E_0$ stands for the interacted energy of the system. With the help of
eqs.~(\ref{e15}) and (\ref{e16}),let us discuss the stability of solutions
shown by eqs.~(\ref{e15}) and (\ref{e20}).
\\
(1) For $r\approx \frac 1{\sqrt 2}$ we have
\begin{eqnarray}
\dot \theta &=& -\frac{J}{4r^2}(2r^2-1)^2 \pm \frac{\omega_0}{r^2}\nonumber,\\
\ddot r &=& -\frac{2r^2-1}{4r^3}\left[ \omega_0(\omega_0\mp J)(2r^2+1)
  +(sr^2-1)^2\right] \equiv F_1(r).
\end{eqnarray}
Since $F_1(\frac 1 {\sqrt 2})=0$
\begin{equation}
F_1(\frac 1 {\sqrt 2}+\delta r) \approx -4\omega_0(\omega_0\mp J)\delta r,
\end{equation}
so that the system is stable if $\omega_0>J$, critical if $\omega_0=J$
and unstable if $\omega_0<J$.\\
(2) For solution given by eq.~(\ref{e16}) we have $r\approx R_{\pm}$, then
\begin{eqnarray}
\hspace*{-2.3cm}
&&\dot \theta = \frac{J}{r^2}(r^2-r^4+\frac{\omega^2_0}4J^{-2})\nonumber,\\
&&\ddot r = \left[ 4J^2(r^2-r^4)-\omega^2_0\right]\frac{4J^2(3r^4-r^2)-
  \omega^2}{16J^2r^3} \equiv F_2(r),\\
&&F_2(R_{\pm}) = 0 \nonumber,\\
&&\hspace*{-1cm}F_2(R_{\pm}+\delta r)
\approx  -8\omega^2_0\left[ \frac{q^4_0}{16\hbar^2
  \omega^2_0}\left(\sum_j\frac{c^2_j}{m_j\omega^2_j}\right)^2-1\right]
  \delta r=-2(\sum_jB_j)^2\delta r.
\end{eqnarray}
Therefore the solution $R_{\pm}$ are stable (with degeneracy).

\section{Semiclassical Equation of Motion}
\setcounter{equation}{0}

To study the dynamic features we follow the standard procedure(see for
example \cite{Ca2}) by assuming N, the number of bath oscillators, is large
enough so that we can replace the sum over $j$ by an integration over
$\omega_i$ in our proceeding discussion. The special distribution in their
path integral approach to the quantum Brownian motion \cite{Ca2}
will be employed. Following
their discussion we shall divide the heat bath effect into three parts, a
renormalization part, dissipative part and random force one and look at how
these terms affect the dynamics of tunnelling and localization.

Let us first give a formal solution of eq.~(\ref{e8}) which takes the form
for each $i$ subscript:
\begin{equation}
B(t)=b(t)+b_0(t)\label{g1}
\end{equation}
where $b_0(t)=c_1\cos  \omega t +c_2\sin \omega t$ and $b(t)$ is a particular
solution which can be expressed by
\begin{equation}
b(t)=c_1(t)e^{i\omega t}+c_2(t)e^{-i\omega t}.
\end{equation}
To find a particular form of $b(t)$ the condition
\begin{equation}
\dot c_1(t)e^{i\omega t}+\dot c_2(t)e^{-i\omega t}=0
\end{equation}
is taken into account. It leads to the well-known solution of eq.~(\ref{g1}):
\begin{eqnarray}
b_0(t) &=& B_j(0)\cos  \omega_jt+\left( \frac{\sin \omega_jt}{\omega_j}\right)
  \dot B_j(0)\label{g2},\\
b(t) &=& -\frac{q^2_0}{2\hbar}\int^t_0S(\tau)\left[\sum_j\frac{c^2_j}
  {m_j\omega_j}\sin \omega_j(t-\tau)\right]d\tau\label{g3}
\end{eqnarray}
or
\begin{equation}
b(t) = -\frac{q^2_0}{\pi\hbar}\int^t_0d\tau S(\tau)\int^\infty_0d\omega
 J(\omega)\sin \omega (t-\tau)
\end{equation}
where
\begin{equation}
J(\omega)=\frac{\pi}2\sum_j\frac{c^2_j}{m_j\omega_j}\delta(\omega-\omega_j).
\end{equation}
Following ref. \cite{Le1}, if the Ohmic approximation is assumed:
\begin{equation}
J(\omega)=A\omega^se^{-\omega/\omega_c},
\end{equation}
then
\begin{eqnarray}
b(t) &=& -\frac{q^2_0A}{\pi\hbar}\int^t_0d\tau S(\tau)\int^\infty_0d
    \omega e^{-\omega/\omega_c}\sin [(t-\tau)\omega]\omega^s \nonumber \\
&=& -\frac{q^2_0A}{\pi\hbar}\omega^{s+2}_c\Gamma(s+2)\int^t_0d
    \tau S(\tau)(t-\tau)\nonumber\\
&&F (\frac{s+2}2,\frac{s+3}2,\frac 32;-\omega^2_c(t-\tau)^2)
\end{eqnarray}
where $F(a,b,c;u)$ is hyper-geometric function.

Let us consider a special case where $s=1$, then
\begin{equation}
\hspace*{-1.4cm}B_0(t)=-\frac{q^2_0A}{\pi\hbar}\omega_cS(t)+\frac{q^2_0A}
  {\hbar}(\frac1{\pi}\frac{\lambda}{\lambda^2+t^2})S(0)+\frac{q^2_0}{\hbar}
  \int^t_0d\tau\frac{dS(\tau)}{d\tau}\left[\frac 1{\pi}\frac{\lambda}
  {\lambda^2+(t-\tau)^2}\right] \label{f2}
\end{equation}
where $\lambda=\omega^{-1}_c$. When $\omega_c\rightarrow \infty$, $A=\eta$,
eq.~(\ref{f2}) is reduced to
\begin{equation}
b(t)=\frac{\eta q^2_0}{\hbar}\frac{dS(t)}{dt}-\frac{\eta\Omega}{\pi\hbar}
     q^2_0S(t).
\end{equation}
This limit is nothing but the Debye distribution by making
$\sum_j\rightarrow \int^{\infty}_{0}d\omega\rho(\omega)$ \cite{Ca2}
\begin{eqnarray}
\rho_D(\omega)\frac{[c(\omega)]^2}{m(\omega)}=\left\{
\begin{array}{ll}2\eta\omega^2 \ &\omega< \Omega\\0&\omega>\Omega\end{array}
\right.\label{g5}
\end{eqnarray}
where $\Omega$ is a high frequency cut-off, $\eta$ is the phenomenalogical
friction coefficient in the Langevin equation eq.~(\ref{e0}).

Let $\Omega\rightarrow\infty$ we have
\begin{equation}
b(t)=\frac{2\eta q_0}{\hbar}\vec r\cdot \dot{\vec r}-\frac{\eta\Omega}{\pi}
\frac{q^2_0}{\hbar}(2\vec r\ ^2-1).
\end{equation}
The second term is much larger than the first one for the usual environment.
In comparison to eq.~(\ref{e9}) we have
\begin{equation}
b(t)\approx -\sum_j\frac{q^2_0c^2_j}{2\hbar m_j\omega^2_j}S(t)
\end{equation}
where
\begin{equation}
\eta\Omega=\pi\sum_j\frac{c^2_j}{2m_j\omega^2_j}.
\end{equation}
With this notation the broken symmetry condition eq.~(\ref{e17})
is rewritten as
\begin{equation}
\frac{\eta\Omega}{2\pi}q^2_0>\hbar \omega_0
\end{equation}
and correspondingly,
\begin{eqnarray}
R_{\pm} &=& \frac{\sqrt{2}}{2}\sqrt{1\pm\sqrt{1-\frac{2\pi^{2}\hbar^{2}
\omega_{0}^{2}}{\Omega^{2}\eta^{2}q_{0}^{4}}}},\\
B_{\pm} &=&
\mp\frac{2}{\hbar}\sqrt{\frac{\Omega^{2}\eta^{2}q_{0}^{4}}
{4\pi^{2}}-4\pi^{2}\hbar^{2}
\omega_{0}^{2}},\\
E_{\mbox{total}}
&=& -\frac{\hbar\omega_{0}}{2}[\frac{\eta q_{0}^{2}\Omega}{2\pi\hbar\omega_{0}}
+\frac{2\pi\hbar\omega_{0}}{\eta q_{0}^{2}\Omega}]\leq-\hbar\omega_{0}
\label{g4}.
\end{eqnarray}
Obviously, eq.~(\ref{g4}) tells that the larger the $\eta\Omega$ is,
the lower the energy of degenerated ground state is.

By virtue of eqs.~(\ref{g1}), (\ref{g2}) and (\ref{g3}), using the continuous
Debye frequency distribution and following \cite{Ca2} we obtain
\begin{eqnarray}
B(t) &=& \int_{0}^{\infty}\rho_{D}(\omega)B(\omega,t)d\omega \nonumber\\
     &=& B_S(t)+F(t)
\end{eqnarray}
where
\begin{equation}
F(t)=\sum B_{i}(0)\cos \omega_{i} t+\sum \frac{\dot{B}_{i}(0)}
 {\omega_{i}}\sin  \omega_{i} t
\end{equation}
is the Langevin force and carries all the characteristics of
random force of classical Brownian motion if we assume the thermal
probability distribution of heat bath as given by eq.~(\ref{g5}).
In fact, it is easy to calculate the average:
\begin{equation}
\langle  \cdots\rangle=\int^{+\infty}_{-\infty}d{\dot
x_j}\int^{+\infty}_{-\infty}dx_j
  \cdots exp\{-\frac{m_j}{2kT}({\dot x_j}^2+\omega^2_jx^2_j)\}
\end{equation}
and find random force:
\[\langle  F(t)\rangle = 0,\]
\begin{equation}
\langle  F(t)F(t')\rangle = 2kT\frac{\eta
q_{0}}{\pi\hbar}\frac{\sin\Omega(t-t')}{t-t'}
\stackrel{\Omega\rightarrow\infty}{\longrightarrow}2kT\frac{\eta q_{0}}
{\hbar}\delta(t-t')
\end{equation}
where $F(t)$ corresponds to $b_0(t)$ in eq.~(\ref{g1}).

With the above knowledge we come to establish the dynamic equation.
We have already known that the angular momentum is related to the
energy of two-level system based on eq.~(\ref{f20}). In
the existence of the interaction with environment the energy is altered
by the magnetic field, namely, the renormalized angular
momentum $L_R$ times $(-\hbar)$ can be viewed as the renormalized energy
$\varepsilon_R$ of the two-level system. The $L_R$ is given by
\begin{equation}
L_{R}(t)=r^{2}\dot{\theta}+\frac{\Omega}{8\pi}\frac{\eta q_{0}^{2}}{\hbar}
S^{2}(t) \;\;\; and \;\;\; \varepsilon_{R}=-\hbar L_{R} \label{g20}.
\end{equation}
Taking the time-derivative of $L_R$ (or $\varepsilon_R$) we get a meaningful
relation similar to the classical Langevin dissipative dynamics:
\begin{equation}
\frac{d}{dt}L_{R}=\frac{\eta q_{0}^{2}}{4\hbar}(\frac{d}{dt}S(t))^{2}
+\frac{1}{4}(\frac{d}{dt}S(t))F(t)\label{g6}\label{g8}.
\end{equation}
The first term on the right hand of eq.~(\ref{g6}) is the dissipative term
which always make $L_{R}$ increase, or in other words, decreased
$\varepsilon_{R}$. The second term represents the work
done by the Langevin force.
In comparison to eq.~(\ref{e0}) $S(t)$ plays the role similar to $q(t)$
so we can expect a Langevin-like dynamic equation for $S(t)$ on the basis
of eqs.~(\ref{e6}) and (\ref{e8}), that can be written in the form:
\begin{eqnarray}
&&\ddot r-r \dot{\theta}^2+\omega^2_0r=-r\dot \theta B(t)\label{g7},\\
&&r \ddot{\theta}+2\dot r\dot{\theta}=\dot r B(t).
\end{eqnarray}
The straight calculation gives
\begin{eqnarray}
\frac{d^{2}}{dt^{2}}S(t)&=&-4\omega_{0}^{2}S(t)+4\alpha L_{R}(t)S(t)-
\frac{1}{2}\alpha^2S^{3}(t)-4(\frac{\eta q_{0}^{2}}{\hbar})L_{R}(t)
\frac{d}{dt}S(t) \nonumber\\
&&+\frac{1}{2}\alpha(\frac{\eta q_{0}}{\hbar})S^{2}(t)\frac{d}{dt}
S(t)-4L_{R}(t)F(t)+\frac{1}{2}\alpha S^{2}(t)F(t)\label{g9}
\end{eqnarray}
where $\alpha=\frac{\Omega\eta}{\pi}\frac{q^2_0}{\hbar}$. In
Comparison with eq.~(\ref{e1}), the
physical meaning of the RHS of eq.~(\ref{g9}) can be explained as follows:

The first three terms are the driven forces, the fourth and fifth terms are
friction force although they do not always make the motion retarded and
the last two terms are random force.

Equations (\ref{g8}), (\ref{g9}) are the central results of our quasiclassical
investigation of dissipative  dynamics of two-level system.
Although the equations are still hard to be solved, a
qualitative consideration can help us insight into some physics of
this quantum coherent dynamics. It is not difficult
to find that the renormalized angular momentum $L_{R}$ plays  important
roles here. It first appears in eq.~(\ref{g20}) representing the renormalized
energy of the system, then in eq.~(\ref{g9}) acting as a time dependent
factor to adjust the forces acting on the charged particle.As we have
studied, for its own properties we
can find that when the broken symmetry condition is not satisfied, $L_{R}$
reaches its minimum value $-\omega_{0}/2$ and the maximum value $\omega/2$.
On the opposite case in which the broken symmetry condition is satisfied
$L_{R}$ still reaches its minimum value $-\omega_{0}/2$ at the eigenstate
(\ref{e15}), but now eigenstate with $L_{R}=\omega_{0}/2$ becomes a
saddle point and $L_{R}$ reaches its maximum value $\frac{\omega_{0}}{4}
[\frac{\Omega}{2\pi}\frac{\eta q_{0}^{2}}
{\hbar \omega_{0}}+\frac{2\pi}{\Omega}\frac{\hbar\omega_{0}}
{\eta q_{0}^{2}}]$ at the true degenerate ground states (\ref{e16}).
Because the dissipation always makes $L_{R}$ increase,
the dynamics with any initial state will tend to ground state
(\ref{e15}) at weak coupling situation and to one of the localized
ground states at strong coupling situation. This is consistent with
our basic knowledge on the broken symmetry picture.

{}From classical mechanics we know that any velocity independent force
$f$ can be written as $f=-dV/dq$. For the first three driven force
terms in (\ref{g9}) we can do the same thing. The explicative potential
can be written in terms of $S$ and $L_R(t)$:
\begin{equation}
V_R(S,t)=\left[2\omega^2_0-2\alpha L_R(t)\right]S^2(t)+\frac 18\alpha^2S^4.
\end{equation}
Here $V_R(S,t)$ is time-dependent and just this time dependence exhibits the
novel dynamic features whether or not the two-level system processes the
broken symmetry. When $\Omega\eta q_{0}^{2}/2\pi
\leq\hbar\omega_{0}$, the time dependent coefficient of $S^{2}$ is never
negative and
$V_{R}(S,t)$ is positive everywhere except at $S=0$, its only minimum point
$V_{R}(S,t)=0$. When $\Omega\eta q_{0}^{2}
/2\pi>\hbar\omega_{0}$ we have a critical value $\hbar\omega_{0}^{2}\pi/
\eta q_{0}^{2}\Omega$ for $L_{R}(t)$. Under the circumstance of $L_{R}(t)\leq
\pi\hbar\omega_{0}^{2}/\Omega\eta q_{0}^{2}$ $V_{R}(S,t)$ has the
similar behavior. As soon as $L_{R}(t)$ exceed this value,
the figure of $V_{R}(S,t)$ will be changed to double well form. At this
situation $S=0$ becomes an unstable equilibrium point and the system at this
point will decay due to external perturbation. At the same time
two new degenerate minimum points appear and they will tend to
$r_{\pm}=\frac{\sqrt 2}2\sqrt{1\pm\sqrt{1-2\pi^{2}\hbar^{2}\omega_{0}^
{2}/\Omega^{2}\eta^{2}q_{0}^{4}}}$ with the increasing of $L_{R}(t)$.
These behaviors of $L_{R}(t)$ agree with the former results of this paper.

Interesting features also occur in the dissipative force terms in (\ref{g9}).
The fifth term, contrary to the ordinary understanding of dissipative
force , behaviors as an advance force and the forth term plays either a
retardative or advance role depending on the positive or negative sign
of $L_{R}(t)$'s value. In other words, the dissipative force may not
only decrease tunneling dynamics but also increase it depending on
the initial condition of the system.

In applying above results, we would like to discuss some special dissipative
dynamic processes in strong coupling limit. Here we only give qualitative
consideration. Detail calculations and comparison with other approaches will
appear in elsewhere.

First we consider the most interesting case with a fully localized initial
wave function, saying $S_{I}(0)=1$. Then from the definition of $L_{R}$
we have $L_{R,I}(0)=\Omega\eta q_{0}^{2}/8\pi\hbar$ and the radial
potential felt by the charged particle likes double well form with
the minimum points very near $\pm 1$. So we can see that the motion can
not leave far away from the edge of the unit circle. Since the velocity
$dS/dt$ is very small, the dissipative force terms in (\ref{g9}) almost do
not affect the motion during any short time period.
The particle moves like a damped oscillator and
losses its kinetic energy with the
change of $L_{R}(t)$. As $L_{R,I}(t)$ reaches its maximum value, the
particle will sit at the minimum point of the potential.

Second let us discuss the case where the eigenstate with $R=\frac{\sqrt 2}2,
\ v=\frac{\sqrt 2}2,\ B=0$ is taken as the initial state.
In this case we can show that $S_{II}(0)=0$, and $L_{R,II}(0)=\omega_{0}/2$.
Now the particle stands on the metastable point and will go out of it
because of thermal fluctuation.

At last suppose the eigenstate with $R=\frac{\sqrt 2}2$, $v=-\frac{\sqrt 2}
2$, $B=0$ is taken as the initial state.
As discussed above we have $S_{III}(0)=0$, and $L_{R,III}(0)=-\omega_{0}/2$.
The radial potential in this case has only one minimum and the particle
is at the equilibrium point initially. Because both the two dissipative
forces are advance ones now, so the range of motion will become larger
and larger in the early period of dynamics. Therefore the particle will
escape this potential well and the shape of potential will be changed into
the double well form

In conclusion we have studied the quasiclassical dissipative dynamics of
a two-level system and  compared it with the ordinary Langevin
description. The further investigation along this direction and fully
quantum mechanical treatment are deserved.

Thanks to  Dr. Jiushu Shao for helpful discussion.
This work was supported in part by the National
Natural Science Foundation of China.

\end{document}